\documentclass[12pt]{elsart}

\usepackage{multirow}
\usepackage{epsfig, setspace, pmgraph}
\usepackage{graphicx}
\usepackage{afterpage}
\usepackage{amsmath, amsfonts, amssymb}
\usepackage{dsfont,courier}
\usepackage{verbatim}
\usepackage{lscape}
\usepackage{longtable}
\usepackage{natbib}
\usepackage{subfig}
\journal{J Mater Process Technol}

\begin{document}

\begin{frontmatter}

\title{Evolution of Plastic Strain During a Flow Forming Process}
\author[cor1]{M. J. Roy\corauthref{cor}}
\ead{majroy@interchange.ubc.ca}
\author{R. J. Klassen\corauthref{cor2}}
\author{J. T. Wood\corauthref{cor2}}

\address[cor1]{Dept. of Materials Engineering, The University of British Columbia, Vancouver, BC, Canada V6T 1Z4}
\address[cor2]{Dept. of Mechanical and Materials Engineering, The University of Western Ontario, London, ON, Canada N6A 5B9}
\corauth[cor]{Corresponding author.}


\begin{abstract}
The distribution of equivalent plastic strain through the thickness
of several AISI 1020 steel plates formed under different conditions
over a smooth cylindrical mandrel using a single-roller forward flow
forming operation was studied by measuring the local
micro-indentation hardness of the deformed material.  The equivalent
plastic strain was higher at the inner and outer surfaces and lowest
at the center of the workpiece. Empirical expressions are presented
which describe the contribution of the roller and mandrel to the
total local equivalent plastic strain within the flow formed part.
The dependence of these expressions upon the thickness reduction
during flow forming is discussed.
\end{abstract}

\end{frontmatter}

\section{Introduction}
In a flow forming operation, a roller (or multiple rollers) are used
to plastically deform the workpiece over the mandrel (Fig. 1).
During this process, the material deforms both axially and
circumferentially thereby simultaneously reducing the thickness and
axially lengthening the work piece.  The flow forming roller can
have a unique profile and this geometry, combined with variable
speeds and feeds, make a simultaneous multivariate analysis of the
plastic strain induced in the workpiece very difficult.
\begin{figure}[]
  \begin{center}
  \includegraphics[width=3 in]{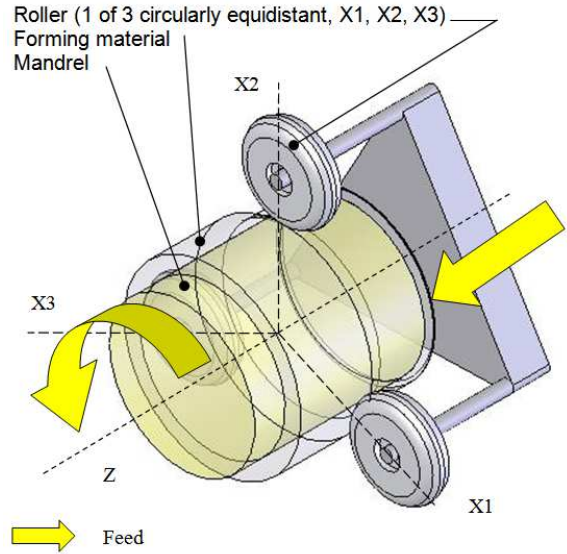}
  \caption[]{Typical three-roller flow forming arrangement.}
  \label{fig:one}
  \end{center}
\end{figure}
\cite{Gur.82} stated that the flow forming process can be thought of
as an operation involving the simultaneous extrusion (or drawing)
and rolling of the workpiece.  Characterization of both how, and the
extent to which, the flow formed material deforms plastically during
this process has not been extensively studied. While \cite{kemin.96}
and \cite{xu.01} have adopted Finite Element Analysis (FEA)
approaches to studying the deformation mechanisms prevalent during
flow forming, there is little experimental work that has been
conducted. As a result, there is a lack of understanding of both
manufacturers and end-use designers of the mechanical properties of
flow formed components, and sources of experimental data with which
to validate FEA results.

The work presented in this paper is aimed at improving the
understanding of the local plastic deformation induced in the
workpiece during a single-roller flow forming process.  While
\cite{brandon.80} studied the strain distribution by inserting pins
into a preform and observing how the pins displaced during forming,
this work is aimed at measuring through-thickness equivalent strain
directly. This is accomplished through experimentally mapping the
local equivalent plastic strain through the thickness of a flow
formed part using micro-indentation hardness measurements.

Micro-indentation techniques have been used by others to infer the
local equivalent plastic strain of highly deformed materials.
\cite{chaudhri.96,chaudhri.00} employed Vickers micro-indentation
techniques to map the local equivalent plastic strain in the
deformed region around a large spherical indentation made in copper
while \cite{tseng.98} also used a similar technique to map the
equivalent plastic strain through the roll-bite region of cold
rolled AISI 1018 steel. Two recent works detailing the
macro-indentation hardness profiles through the thickness of flow
formed ferritic steel workpiece alloys were published by
\cite{chen&jones.02} and \cite{gur.03}, however neither of these
investigations correlated the measured hardness to the equivalent
plastic strain.

\section{Experiment Details}\label{sec:exp6details}

Single-roller flow forming operations were performed over a smooth
cylindrical mandrel using a roller with an outer surface profile
consisting of two flat regions and a blending radius between them
(Fig. 2).  Three passes (Fig. 3, 4) of the roller were used to form
flat circular blanks of AISI 1020 steel plate, of 8.5 mm initial
thickness, over the mandrel. During the first pass, the part did not
contact the mandrel; this is essentially a classical metal spinning
operation. The second pass involved minimal contact of the workpiece
with the mandrel. The work piece was in full contact with the
mandrel throughout the third pass. The forming process was stopped
in the middle of the third pass by quickly retracting the roller
away from the work piece.  This was done to investigate the local
deformation of the steel in the region ahead of the roller. In
total, six tests were performed.  The first two forming passes of
each test were performed under the same conditions. The thickness
reduction for the first pass was 6\% and the second pass was 29\%,
while the third pass was performed at six different thickness
reduction levels, resulting in a total thickness reduction ranging
from 48.2 to 55.3\%.

\begin{figure}[]
  \begin{center}
  \includegraphics[width=6 in]{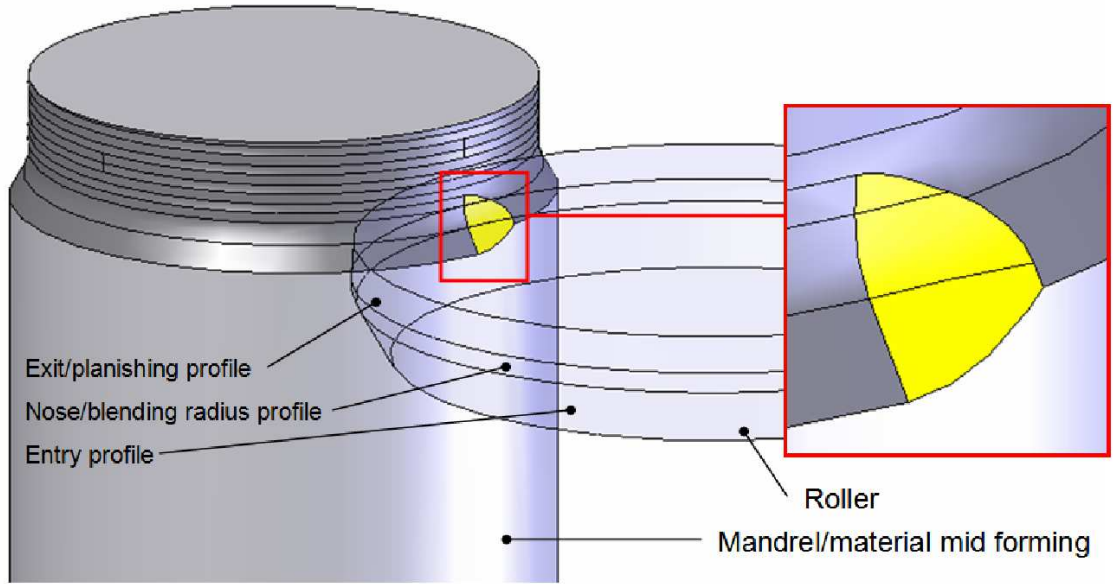}
  \caption[]{Single roller contact in flow forming showing the mandrel/workpiece and key roller profiles.}
  \label{fig:two}
  \end{center}
\end{figure}

\begin{figure}[]
  \begin{center}
  \includegraphics[width=6 in]{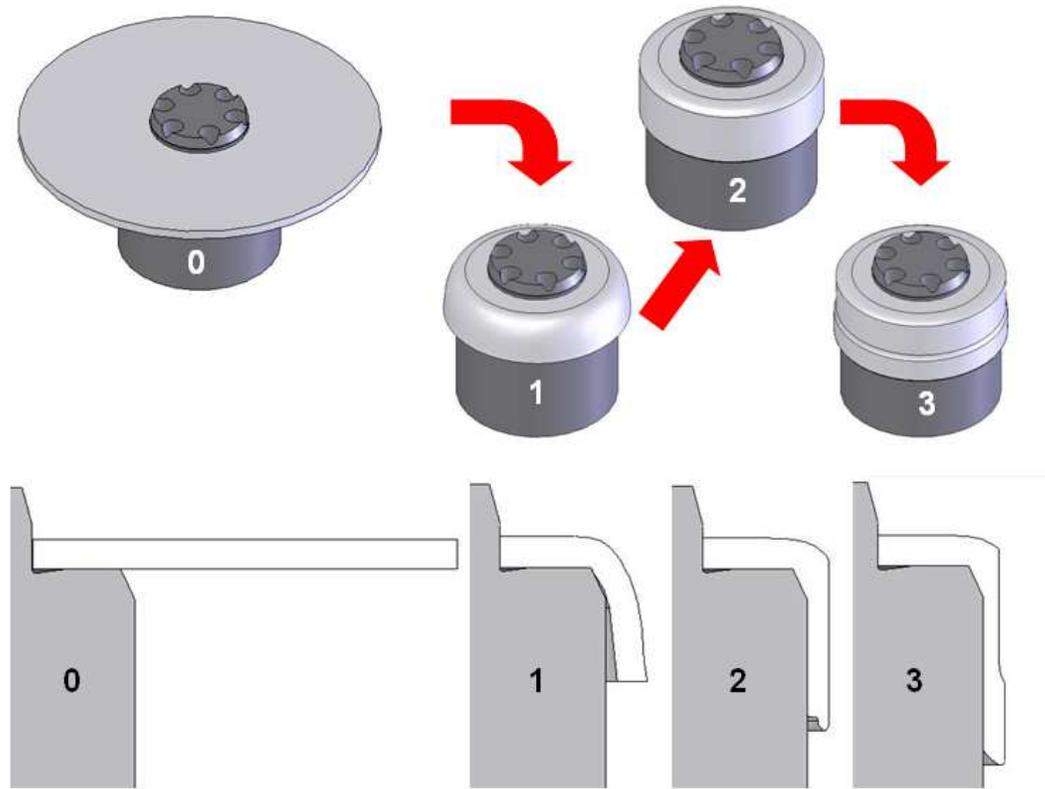}
  \caption[]{Flow forming sequence showing (0) starting
                  material, (1) `cupped' material, (2) first forming pass and (3) later pass
                  frozen mid-forming.}
  \label{fig:three}
  \end{center}
\end{figure}

\begin{figure}[]
  \begin{center}
  \includegraphics[width=6 in]{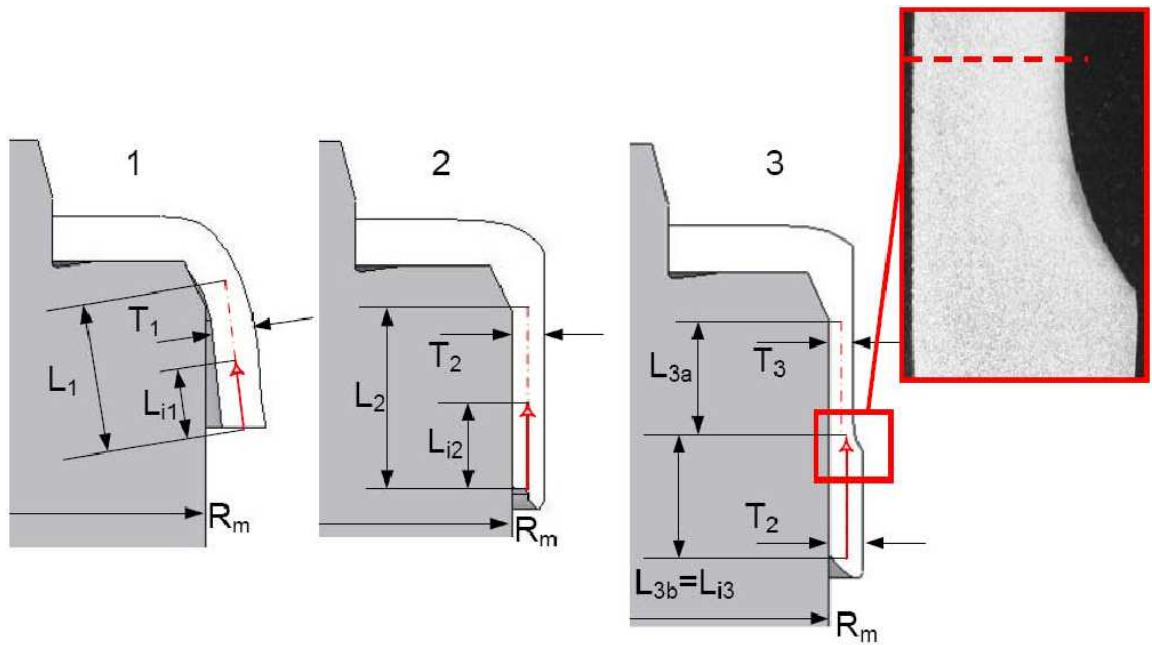}
  \caption[]{Indentation location at distance $L_i$ from the base of each of the three successive parts shown in
                  Fig. 3.}
  \label{fig:one}
  \end{center}
\end{figure}

Each flow formed part was inspected to find the final roller
position on the part. The part was then sectioned at this point and
mechanically polished to $<1\mu$m surface roughness in preparation
for micro-indentation hardness testing.  Depth controlled
micro-indentation hardness data were collected from the polished
as-received 1020 steel and from the flow formed parts using a
Berkovich microindentation hardness tester manufactured by Micro
Materials Ltd. (Wrexham, UK).  It was found that the optimal
indentation spacing was 100 $\mu$m, and the optimal indentation
depth was  $\sim 7\mu$m.

Three rows of indentations were made on each of the six samples.
Since the objective of the indentation testing was to characterize
the evolution of the local plastic strain at a specific location
within the workpiece and since the workpiece was incrementally
deformed through multiple flow forming passes, the specific
location, $L_i$ of the indentations (Fig. 4), must be different for
each successive forming pass. This was done to guarantee that the
indentations are made in the same location within the workpiece.
Assuming that work piece volume is conserved during the deformation,
and using the known dimensions $T_1$, $T_2$, $T_3$, $L_{3a}$ and the
mandrel radius, $R_m$, the location of the central rows of indents
$L_{i1}$ and $L_{i2}$ were calculated (Fig. 4). This assumes that
the work piece geometry evolves through each of the forming passes
as a hollow cylinder becoming both thinner and longer.

\section{Correlating Micro-Indentation Hardness to Equivalent Plastic Strain}
The correlation of the Berkovich hardness values to equivalent
plastic strain was accomplished by deforming  as-received AISI 1020
steel plate to different levels of plastic strain through
cold-rolling and uniaxial tensile deformation.  Seven specimens were
treated in this manner with true plastic strain from rolling ranging
from 0.25 to 0.40. For small strain levels, five tensile specimens
were cut directly from as received material. The tensile tests were
conducted on a Instron 8804 servo-hydraulic test platform according
to standard procedures (ASTM 370), with specimens having a gauge
length of 50.8 mm. For all tensile tests, the extension rate was set
at 2 mm per minute, and the specimen extension was continually
monitored with a clip-on type extensiometer.

The tensile specimens were then sectioned and polished. The same
micro-indentation procedure was then performed on these samples as
was performed on the flow formed samples.  Rows of indentations were
made in locations through the thickness, outside of the necked
regions of the sectioned tensile samples, on surfaces whose normal
was perpendicular to the rolling direction. A minimum of 38 indents
up to a maximum of 50 indents per sample were made to render
statistically significant results in locations that such that all
indents were made far enough away from the edges of the tensile
specimen that they were not affected by surface/corner inhomogeneity
of the plastic strain. Fig. 5(a) shows the measured Berkovich
hardness plotted against the equivalent plastic strain for both the
cold-rolled specimens and the uniaxially deformed tensile specimens.
\begin{figure}
\centering
  \subfloat[]{\label{fig:5a}\includegraphics[width=3 in]{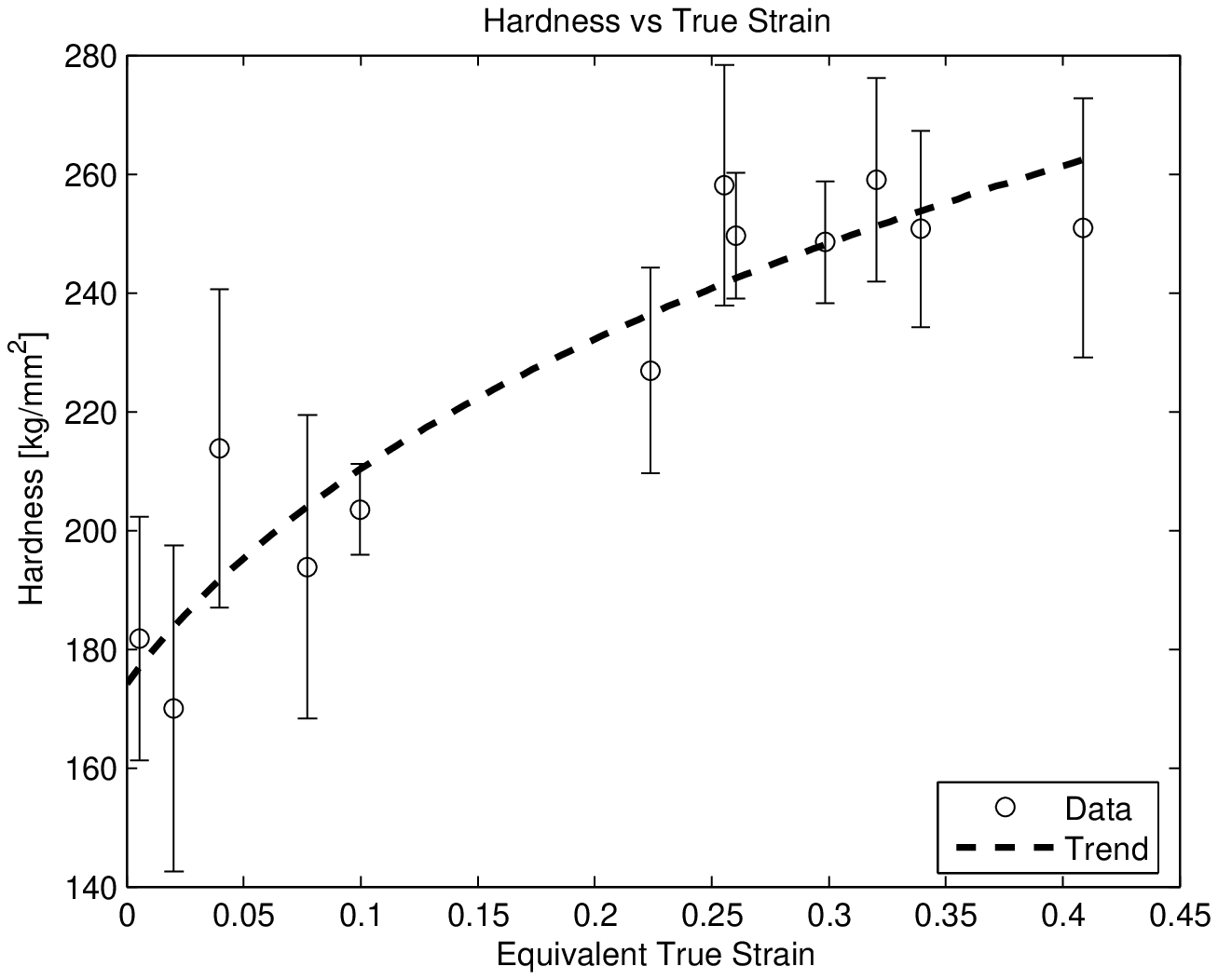}}
  \subfloat[]{\label{fig:5b}\includegraphics[width=3 in]{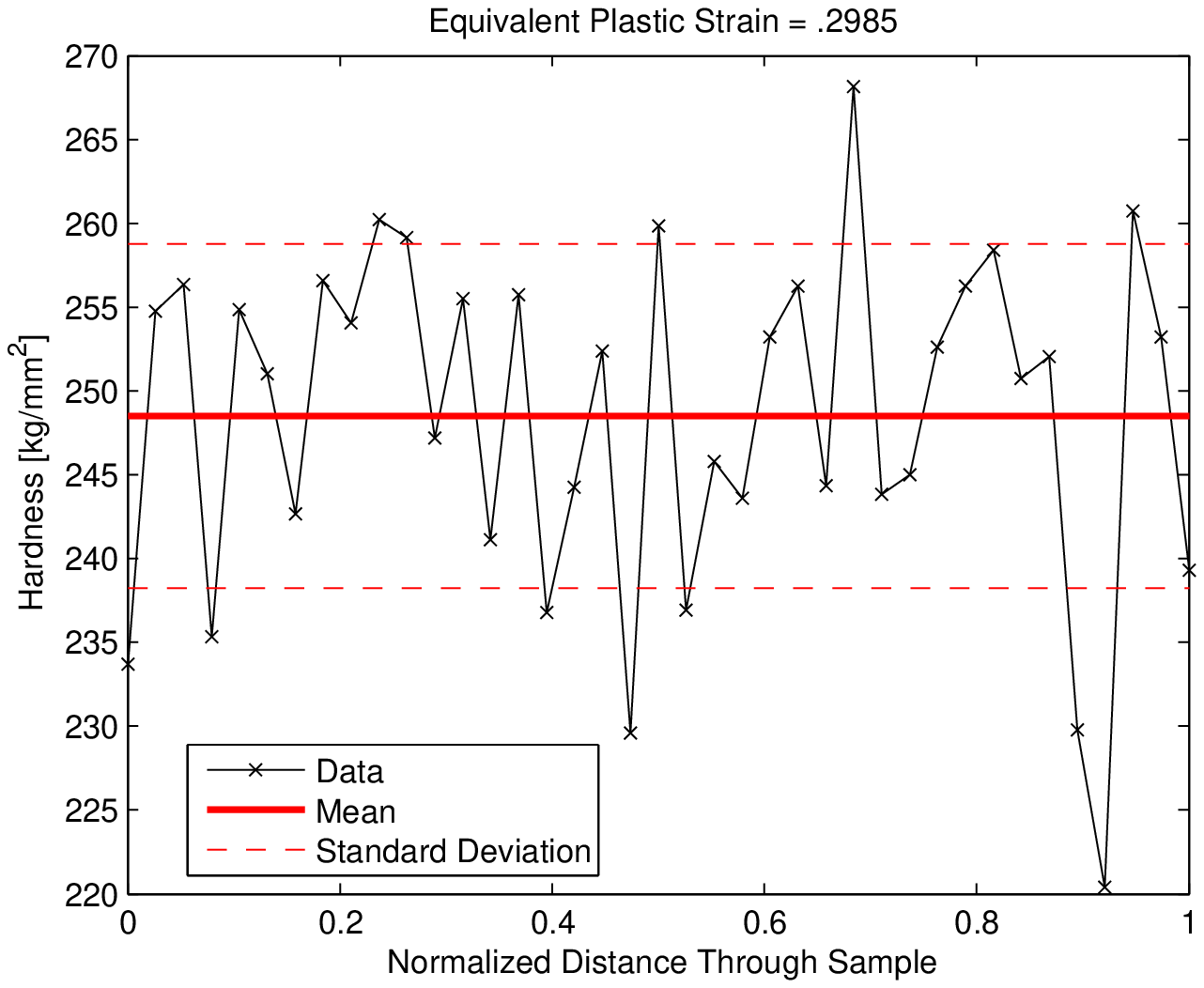}}\\
  \subfloat[]{\label{fig:5c}\includegraphics[width=3 in]{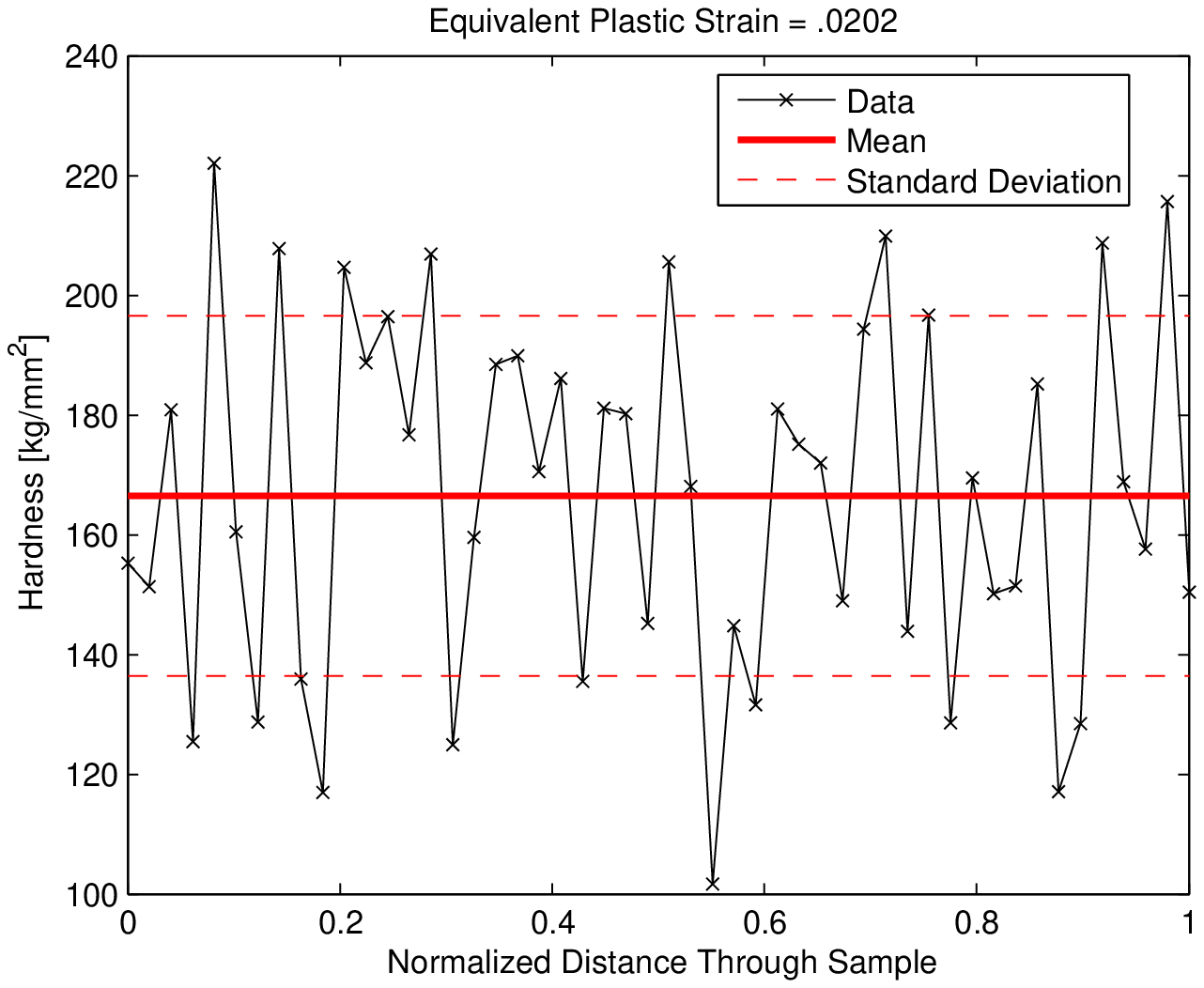}}
  \caption{(a)Indentation hardness versus true equivalent plastic strain of
                  AISI 1020 steel tensile samples cold rolled to various levels of
                  equivalent plastic strain prior to tensile testing.  Error bars are
                  $\pm$1 standard deviation from mean hardness values, and show
                  inherent scatter from $\pm$ 4\% at high strains to $\pm$ 16\% at low
                  strains. The sample with both the highest scatter at a strain of
                  0.0202 (b) and the lowest scatter at a strain of 0.2985 (c) are
                  shown with hardness values plotted versus normalized distance
                  through the thickness of the sample.}
  \label{fig:five}
\end{figure}
As the micro-indentation scheme selected was one which minimized
indentation depth to minimize the spacing between the indentations,
the effect of local mechanical anisotropy of the steel appears in
the form of scatter in the measured hardness. The error bars on the
average hardness data in Fig. 5 indicate one standard deviation of
variance in the measured hardness about the calculated mean hardness
value. While the average percentage variation in the hardness is
about 8\% at any given value of equivalent plastic strain, this
percentage rises to 16\% when the plastic strain is low. For the
highly deformed rolled specimens, the scatter is diminished and
about 4\%. This scatter is inherent for this type of measurement, as
the measured indentation hardness is an average of the yield
properties of all the grains that are deformed in the immediate
region around the indentation. Since the rolling process compresses
the grain size through the thickness of the sample normal to the
rolling direction, indentations made in the rolled material interact
with more grains resulting in a measured hardness that displays less
scatter from indentation-to-indentation. Figures 5(b) and (c) show
the hardness plotted across the normalized sample thickness for both
high and low strain samples.

Following the work of \cite{tabor.51}, it was found that the
correlation of the Berkovich microhardness with the equivalent
plastic strain, $\varepsilon$, can be fitted with the following
expression.

\begin{equation}\label{eq:ch6generalmodel}
H(\varepsilon) = A\left( {\varepsilon  + \varepsilon _{ind} } \right)^n
\end{equation}

In this equation, $H$ is the Berkovich hardness in kg$_f$/mm$^2$ at
depth of $\sim$ 7 $\mu$m, $A$ and $n$ are material constants and
$\varepsilon _{ind}$ is the additional average equivalent plastic
strain associated with the indentation process. These parameters
were all derived from curve-fitting to the data in Fig. 5. The trend
acquired in this manner described the experimental results within
$\pm$1 standard deviation with an $R^2$ value of 0.85 ($R^2$ value
of 1 being a perfect fit) and a Standard Square Error of 1.74E03.
Rearranging Eq. \ref{eq:ch6generalmodel} and substituting the fitted
terms $A$, $n$ and $\varepsilon_{ind}$ gives the following expression
for the average equivalent plastic strain as a function of hardness.

\begin{equation}\label{eq:generalmodel}
\varepsilon(H)  = \left( {\frac{{H }}{{307.1}}} \right)^{4.673}  -
0.0707
\end{equation}

The fitted term $\varepsilon_{ind}=0.0707$ is close to
$\varepsilon_{ind}=0.08\%$ reported by \cite{tabor.51} for pyramidal
indentations made in low carbon steel.

\section{Results}

The hardness profiles from the flow formed samples were translated
to equivalent true plastic strain using Eq. \ref{eq:generalmodel}.
Fig. 6 shows the results of this translation for samples deformed
with one to three flow forming passes.

\begin{figure}[]
  \begin{center}
  \includegraphics[width=3 in]{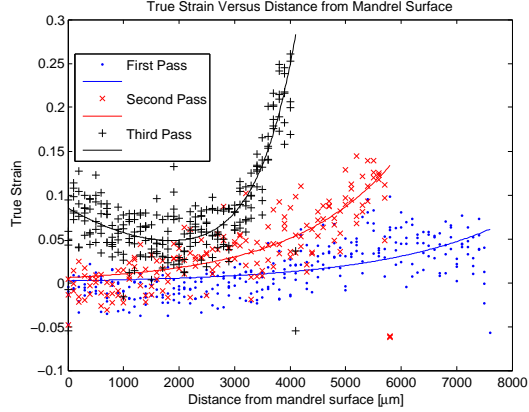}
  \caption[]{True equivalent plastic strain found from Berkovich micro-hardness testing versus distance from mandrel surface for three forming passes resulting in a 50.5\% total reduction.}
  \label{fig:six}
  \end{center}
\end{figure}

For the first pass, the work piece is only acted on by the roller;
the mandrel does not contact the workpiece.  The equivalent plastic
strain at the mandrel surface is therefore approximately zero (Fig.
6). During the second pass there is minimal contact between the
workpiece and the mandrel, the equivalent plastic strain is close to
zero at the mandrel surface and increases up to about 15\% toward
the roller side of the work piece. During the third pass, the work
piece is in full contact with the mandrel and there is a marked
increase, up to about 10 \%, in equivalent plastic strain at the
mandrel surface. The plastic strain at the roller side of the work
piece is much greater than the previous pass due the large
additional plastic strain imparted in the third pass.

While micro-indentation is a robust way of measuring local
equivalent plastic strain, and has been used for this purpose by
other researchers, the magnitude of the local plastic strain is also
reflected in local changes in grain size and shape in the deformed
part. Scanning electron microscopy (SEM) was therefore used to
record the local grain shape around indentations in order to confirm
that the variation in local equivalent plastic strain, measured by
the micro-hardness technique, coincided with the observed change in
local grain size and shape.  The SEM analyses were performed on
indented, third pass flow formed samples that were etched in a 10\%
nital solution for 5 seconds after the indentations were made (Fig.
7).

At the roller surface, the grains are much more refined in the
thickness direction and elongated in the axial direction than at the
minimum strain location in the center of the material. The grains at
the mandrel surface are more refined than at the minimum strain
location, but less than at the roller surface. This change in grain
morphology agrees with the known nature of flow forming deformation
as published by \cite{chen&jones.02} and \cite{gur.03}; namely the
rollers compress the material in the thickness direction while
elongating it in the axial direction. Also apparent in Fig. 7 is the
relative size of the indentations; smaller indentations correlate to
higher degrees of grain refinement and elongation.  The
microstructure observations therefore corresponds to the trends
shown in Fig. 6.

\begin{figure}
\centering
  \subfloat[]{\label{fig:7a}\includegraphics[width=3 in]{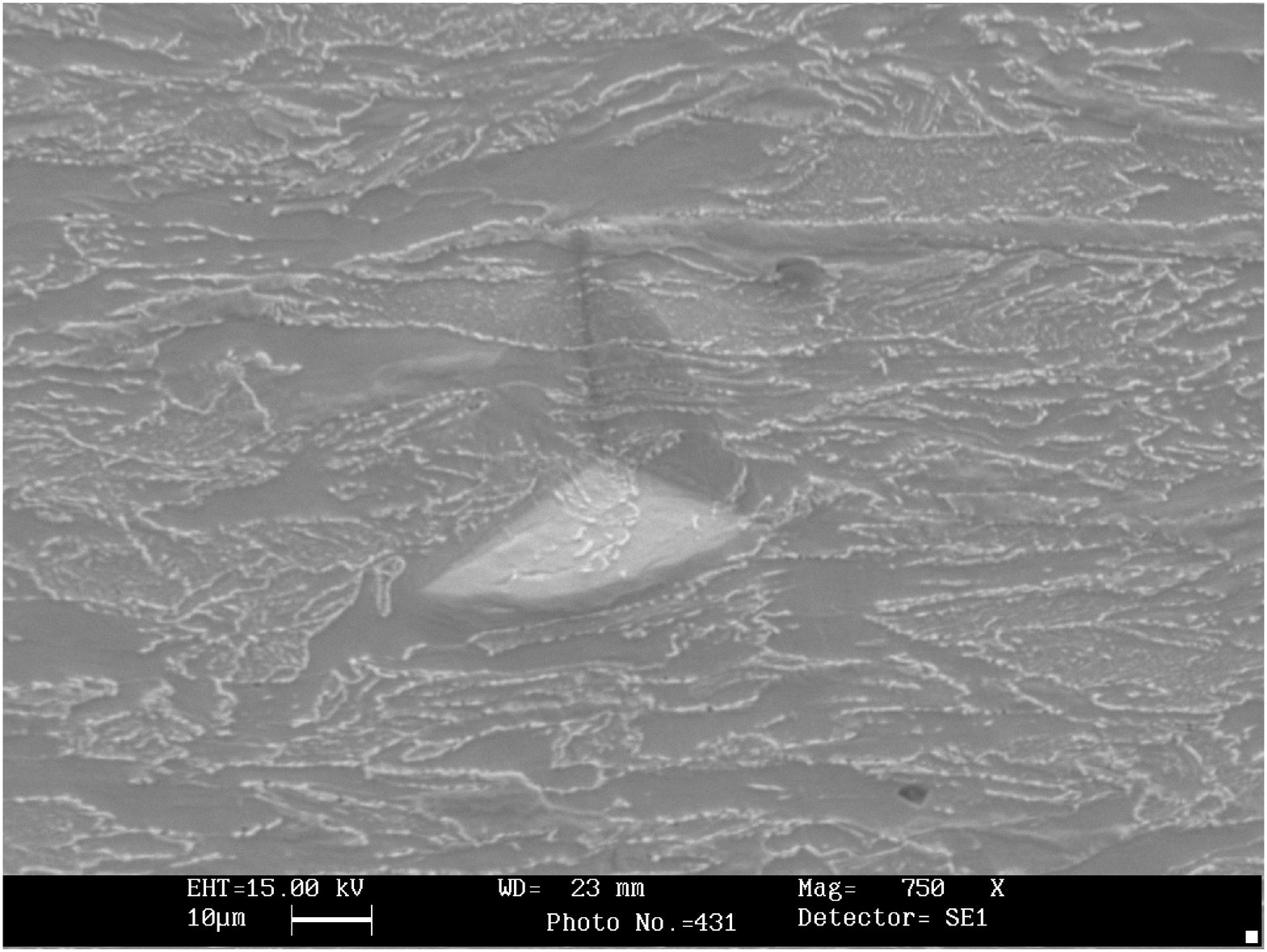}}
  \subfloat[]{\label{fig:7b}\includegraphics[width=3 in]{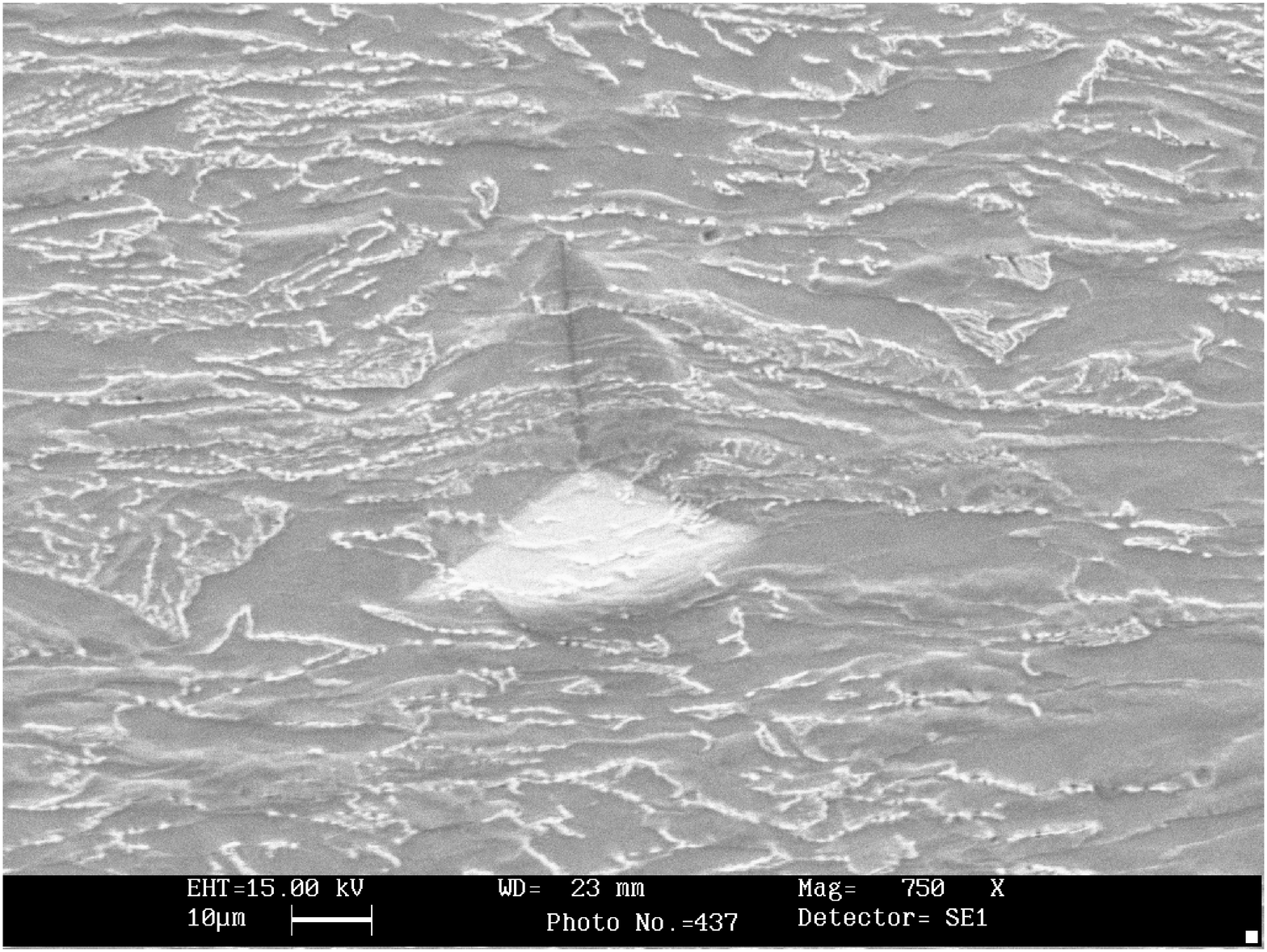}}\\
  \subfloat[]{\label{fig:7c}\includegraphics[width=3 in]{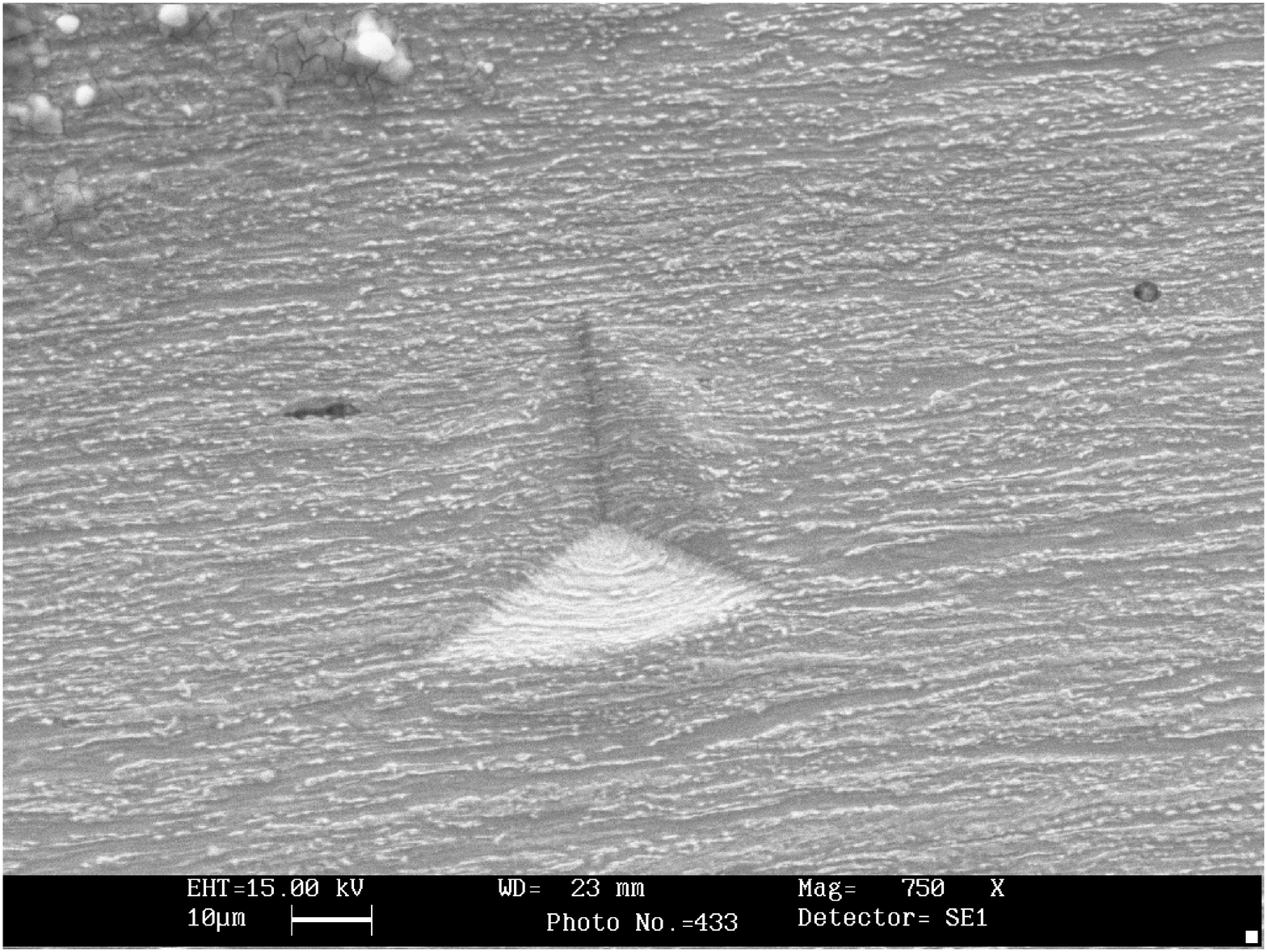}}
  \caption{SEM images of Berkovich indentations through at (a) close to the mandrel interface, (b) minimum hardness and (c) close to the roller interface of a flow formed work piece.
                  Local grain elongation is apparent, as is small decrease in the relative size of the indentations in areas where the grain elongation is largest.}
  \label{fig:seven}
\end{figure}

\section{The Effect of the Roller and Mandrel on the Local Equivalent Plastic Strain}\label{sec:analysis}

The data in Fig. 6 can be used to derive expressions relating the
evolution of the equivalent true strain across the thickness of a
flow formed AISI 1020 steel work piece.  This is accomplished by
plotting the equivalent plastic strain versus normalized thickness
across the work piece and expressing the profiles in the form of
exponential relationships (Eq. \ref{eq:total}-\ref{eq:roller}). This
allows for the best possible expression while minimizing the number
of empirical terms to describe the effects of both the roller and
mandrel's influence on the material.

For the first pass, plastic strain due to the influence of the
roller, $\varepsilon _{1r} (x)$ is imparted to the material, as shown
in Eq. \ref{eq:roller}, where $x$ is the normalized distance from
the mandrel. For the second pass, the strain developed in the first
pass is added to by the second pass of the roller, $\varepsilon _{2r}
(x)$. It should be noted that it is assumed that there is minimal
contact of the workpiece with the mandrel during the first and
second roller passes, and as such $\varepsilon _{1m} (x)=0$ and
$\varepsilon _{2m} (x)=0$. This assumption is validated by the
experimental data in Fig. 6 that show essentially no equivalent
plastic strain at the mandrel surface for these passes. For the
third pass, the workpiece is in complete contact with both the
roller and the mandrel. Therefore, the equivalent plastic strain is
added to by contributions from both the roller and the mandrel,
$\varepsilon _{3r} (x)$ and $\varepsilon _{3m} (x)$.

\begin{gather}
\varepsilon (x) = \sum\limits_{i = 0}^n {\varepsilon _i } (x)\label{eq:total}\\
\varepsilon _i (x) = \varepsilon _{im} (x) + \varepsilon _{ir} (x)\label{eq:each}\\
\varepsilon _{im} (x) = \frac{{A_{im} \Delta t}}{{t_o }}\exp \left(
{\frac{{B_{im} x}}{{t_o - \Delta t}}} \right)\label{eq:mandrel}\\
\varepsilon _{ir} (x) = \frac{{A_{ir} \Delta t}}{{t_o }}\exp \left(
{\frac{{B_{ir} x}}{{t_o - \Delta t}}} \right)\label{eq:roller}
\end{gather}

Where $x$ is the normalized thickness, $\varepsilon(x)$ is the
equivalent plastic strain as a function of the normalized thickness,
$n$ is the total number of forming passes, $i$ is each forming step.
$\varepsilon2_i(x)$ is the equivalent plastic strain occurring at each
forming step and can be furthermore expressed as the sum of
$\varepsilon_{im}(x)$, the equivalent plastic strain due to the
mandrel, and $\varepsilon_{ir}(x)$, the equivalent plastic strain due
to the roller at each forming step. For the formulations of
$\varepsilon_{im}(x)$ and $\varepsilon_{im}(x)$ (Eqns. 5 and 6),
$A_{im,ir}$ and $B_{im,ir}$ are fitted empirical coefficients, while
$\Delta t$ and $t_o$ are the change in thickness and the starting
thickness, respectively.

Through normalization, the coefficients achieved through fitting,
$A_{im,ir}$ and $B_{im,ir}$ (Table \ref{table:results}) are
dependent on process conditions and the length along the part,
$L_i$, and not strain. Note that the coefficients used for the first
and second forming pass were derived from the first and second pass
data from the sample with 55.3\% reduction.

\begin{table}[htbp]
\caption{Fitted coefficients to Eq. \ref{eq:total} at various
reduction levels with confidence boundaries and Sum of Squared Error
(SSE).  Third pass results are averaged over all six
samples.}\label{table:results}
\begin{singlespacing}
\begin{center}
\begin{tabular}{cccc}
 \hline
\multirow{3}{*}{ Sample} & \multirow{3}{*}{Coefficient}&Value & \multirow{3}{*}{SSE}\\
&&(95\% confidence  & \\
&&bounds)& \\
 \hline
\multirow{2}{*}{First pass}&$A_{1r}$&4.10E-01 ($\pm$1.27E-03) & \multirow{2}{*}{0.1634}\\
&$B_{1r}$ &4.20 ($\pm$1.11E-01) & \\
\hline
\multirow{2}{*}{Second pass}& $A_{2r}$ &4.20E-03 ($\pm$1.96E-03) & \multirow{2}{*}{0.1148}\\
&$B_{2r}$ &1.60 ($\pm$3.05E-01) & \\
\hline
\multirow{4}{*}{Third pass}& $A_{3m}$ &1.90E-01 ($\pm$2.75E-02) & \multirow{4}{*}{0.3711}\\
&$B_{3m}$ &-1.31 ($\pm$6.52E-01) & \\
&$A_{3r}$ &5.54E-04 ($\pm$6.23E-04) & \\
&$B_{3r}$ &8.06 ($\pm$1.84) & \\
\hline
\end{tabular}
\end{center}
\end{singlespacing}
\end{table}

Each of the terms in Eq. \ref{eq:total} are a function of process
parameters such as roller geometry, mandrel geometry,  the rate of
axial roller movement, the mandrel rotational velocity and the
degree of thickness reduction per pass. In this investigation, only
the effect of thickness reduction ratio during the third and final
pass is assessed while all other variables are held constant. Fig. 8
shows $\varepsilon _{3r} (x)$ and $\varepsilon _{3m} (x)$ versus
through-thickness distance, $x$, from the mandrel surface for the
thickness reductions ranging from 48.2\% to 55.2\%.  Note that the
relationships derived here are based upon data obtained by using
Eqn. 2, and therefore are subject to an additional average scatter
of about $\pm$ 8\% resulting in the measured indentation hardness
for samples containing a fixed value of average equivalent plastic
strain (Section 3).

\begin{figure}
\centering
  \subfloat[]{\label{fig:8a}\includegraphics[width=3 in]{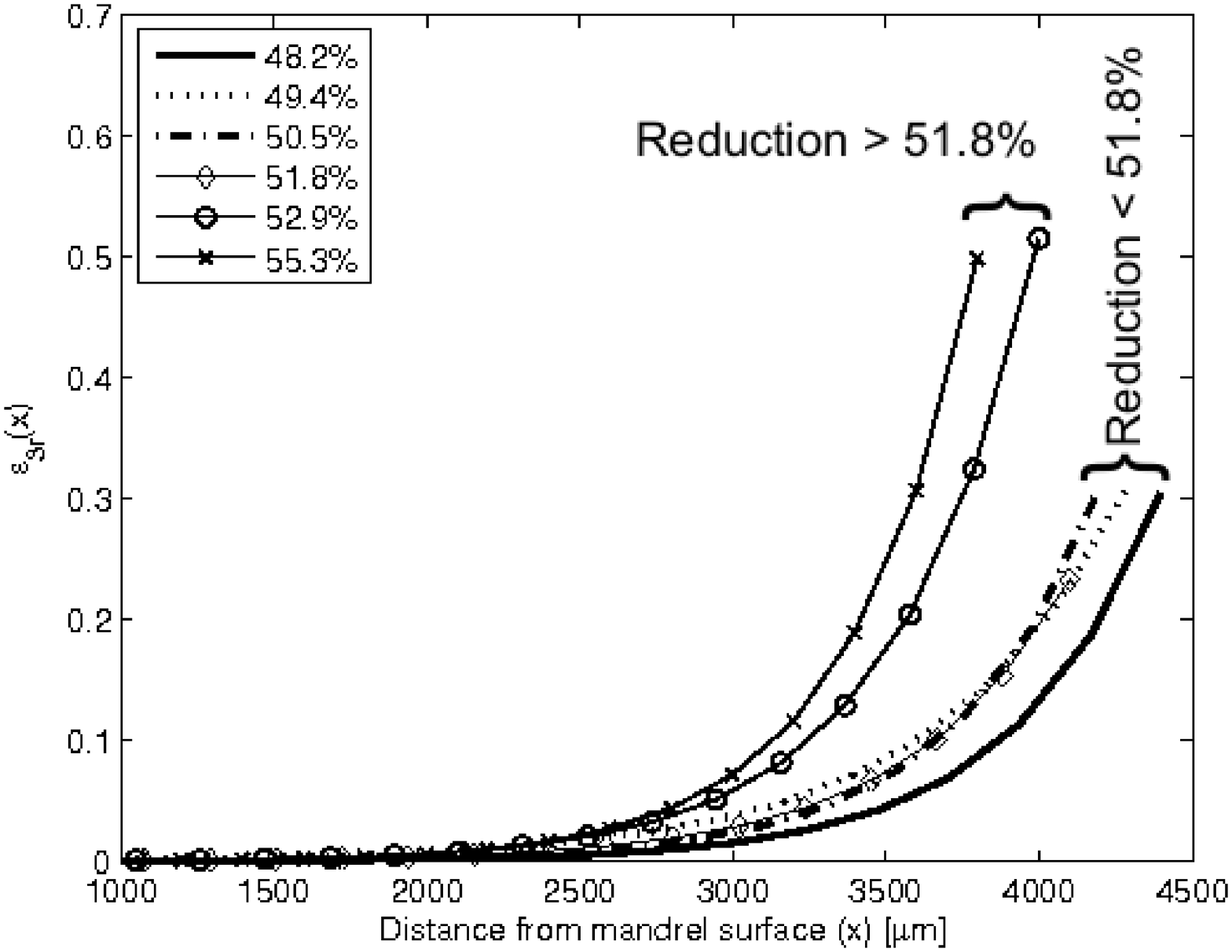}}
  \subfloat[]{\label{fig:8b}\includegraphics[width=3 in]{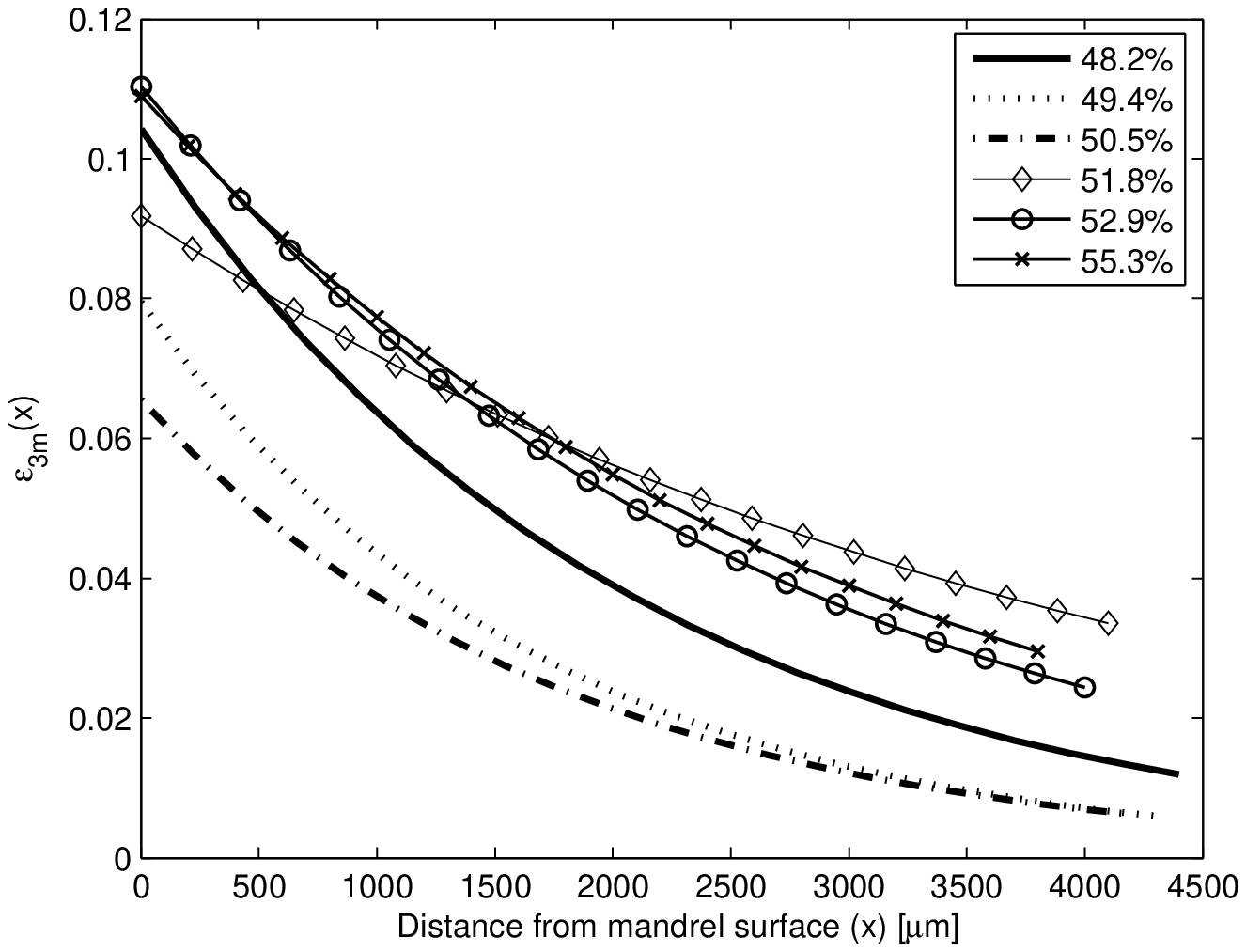}}\\
  \caption{Equivalent plastic strain trends through the thickness of a
                  flow formed workpiece at three different reduction levels. $\varepsilon
                  _{3r} (x)$, shown in (a) is a function due to the roller, and $\varepsilon _{3m}
                  (x)$, shown in (b),
                  is a function due to the mandrel.  A critical reduction level past
                  51.8\% results in a large increase in $\varepsilon _{3r} (x)$.}
  \label{fig:eight}
\end{figure}

For the thickness reduction levels investigated, the shape of the
trend $\varepsilon _{3m} (x)$ essentially stays constant.  For the
roller side, the change in the equivalent plastic strain versus
position profile is more dramatic with the gradient increasing as
the thickness reduction is increased.  The presence of extreme
strain gradients in forming operations often leads to defects in the
final product.  The presence of large strain gradients after a
critical reduction level suggests that there is a maximum thickness
reduction level at which the material can be flow formed and still
remain defect free. This was found to occur at a reduction level
between 51.8\% and 52.9\%.  The overall distribution of strain
changes abruptly, as shown in Fig. 8(b), leading to an increase in
maximum strain of 0.2 at the roller interface.  This is shown in
Fig. 9, where the maximum strain at the roller interface versus
reduction level shows an abrupt shift as the thickness reduction
level passed between these two values.  At reduction values beyond
51.8\%, defects, in the form of localized cracking on the
roller-side surface of the formed part were observed. As indentation
testing was conducted away from these defects, the presence of these
defects do not affect the findings of this paper.   The size, scale
and evolution of these defects remain the subject of on-going
research.

\begin{figure}[]
  \begin{center}
  \includegraphics[width=3 in]{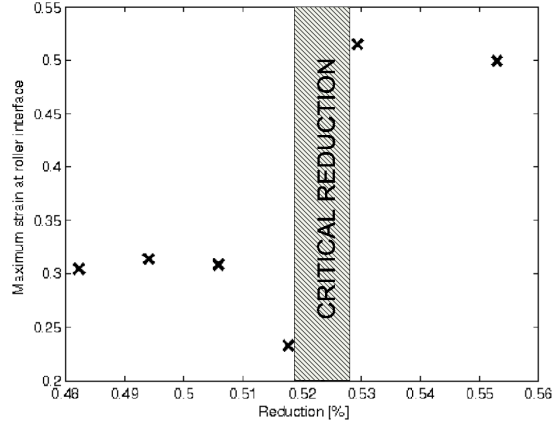}
  \caption[]{Maximum equivalent plastic strain incurred at the roller interface found from fitted relationships versus
                  thickness reduction level.  The strain increases substantially at
                  a critical reduction level between 51.8\% and 52.9\% indicated by the vertical bar.}
  \label{fig:nine}
  \end{center}
\end{figure}

\section{Conclusions}
Micro-indentation hardness testing was successfully employed to map
the true equivalent plastic strain through the thickness of a flow
formed AISI 1020 steel work piece that was deformed by a three-stage
single roller, forward flow forming process.  The work piece
experienced increased plastic strain in subsequent forming passes
with material near the mandrel and the roller displaying elevated
equivalent plastic strain, which was dependent upon thickness
reduction, during the final forming stage. This coincided with the
onset of complete contact between the work piece and the mandrel. It
was also observed that as reduction increased, the local plastic
strain increased more rapidly at the roller interface than at the
mandrel interface.  This trend increased very rapidly past a
critical reduction level found to be between 51.8 and 52.9\%.
Therefore, it is suggest that since there is a substantial increase
in plastic strain at the roller interface when the thickness
reduction is about 52\%, this represents the maximum equivalent
plastic strain that can be imparted to the 1020 steel by flow
forming prior to the onset of roller-induced defects on the surface
of the workpiece. This characterization of roller-induced defects
created during high-strain flow forming is the subject of further
investigation.

In summary, this investigation has presented a characterization of
the effect of the roller, the mandrel and thickness reduction
process parameter on the evolution of the local equivalent plastic
strain in a work piece during flow forming.  The approach and
results of this investigation allows to validate further FEA
modeling approaches as well as for end users of flow formed
components to understand and quantify the mechanical properties of
flow formed parts.

\newpage

\end{document}